Building a Standardised Statistical Reporting Toolbox in an Academic Oncology Clinical Trials Unit: The grstat R Package

Dan CHALTIEL[1,2], Alexis COCHARD[1,2], Nusaibah IBRAHIMI[1,2], Charlotte BARGAIN[1,2], Ikram BENCHARA[1,2], Anne LOURDESSAMY[1,2], Aldéric FRASLIN[1,2], Matthieu TEXIER[1,2], Livia PIEROTTI[1,2]

1 Gustave Roussy, Office of Biostatistics and Epidemiology, Université Paris-Saclay, Villejuif, France
2 Inserm, Université Paris-Saclay, CESP U1018, Oncostat, labeled Ligue Contre le Cancer, Villejuif, France

## Abstract

Academic Clinical Trial Units frequently face fragmented statistical workflows, leading to duplicated effort, limited collaboration, and inconsistent analytical practices. To address these challenges within an oncology Clinical Trial Unit, we developed **grstat**, an R package providing a standardised set of tools for routine statistical analyses.

Beyond the software itself, the development of **grstat** is embedded in a structured organisational framework combining formal request tracking, peer-reviewed development, automated testing, and staged validation of new functionalities. The package is intentionally opinionated, reflecting shared practices agreed upon within the unit, and evolves through iterative use in real-world projects. Its development as an open-source project on GitHub supports transparent workflows, collective code ownership, and traceable decision-making.

While primarily designed for internal use, this work illustrates a transferable approach to organising, validating, and maintaining a shared analytical toolbox in an academic setting. By coupling technical implementation with governance and validation principles, **grstat** supports efficiency, reproducibility, and long-term maintainability of biostatistical workflows, and may serve as a source of inspiration for other Clinical Trial Units facing similar organisational challenges.

## Background and motivation

In many academic Clinical Trial Units, statisticians typically work in parallel, each developing their own scripts, workflows, and conventions. While this autonomy can be effective in the short term, it leads to a number of challenges: duplicated effort, inconsistent practices, and a steep learning curve for new staff, who must navigate diverse codebases and documentation.

In our Clinical Trial Unit, a suite of analysis macros implemented in SAS statistical software has long helped mitigate these challenges by providing some level of standardisation across projects. However, the adoption of R has expanded in recent years, motivated both by cost efficiency and recruitment constraints, as most universities no longer teach SAS. This evolution offered an opportunity to rebuild these tools while also broadening their scope and incorporating modern programming practices.

Together, these considerations highlighted the need for a tool capable of supporting routine statistical analyses while promoting consistent practices across projects. Such a tool needed to be flexible enough to accommodate diverse study designs, yet structured enough to facilitate code reuse, reproducibility, and onboarding within the unit.

## Development of a standardised analytical toolbox for clinical trials

To address this need, a working group was established within the Clinical Trial Unit, bringing together statisticians with complementary expertise in methodology and programming. The group focused on defining common analytical standards and implementing them through a shared toolbox, developed iteratively and aligned with existing workflows. Particular attention was given to structural aspects, including code review, documentation, and testing, to ensure robustness and long-term maintainability.

While the resulting tool addresses needs specific to our Clinical Trial Unit, the underlying development process and governance model were designed to be generic. In particular, the combination of shared standards, structured contribution workflows, and staged validation could be adapted to other clinical research units facing similar challenges, independently of the specific analytical implementations.

## The grstat R package: structure and key features

The outcome of this initiative is **grstat**, an R package primarily designed for internal use within the Clinical Trial Unit. An R package provides a structured framework for bundling functions and documentation into a single, reusable tool, making it a natural solution for supporting shared and maintainable statistical workflows. By adopting an open-source development model on GitHub (https://github.com/Oncostat/grstat), the project aligns with open science principles of transparency.

The package is intentionally opinionated, reflecting shared practices agreed upon within the unit. It provides a structured collection of core functions covering routine statistical tasks commonly encountered in clinical trial analyses, alongside utility functions intended to support reproducible workflows. Its design emphasises consistency of outputs, modularity, and ease of reuse across projects.

The package includes detailed documentation provided at two complementary levels. Function-level documentation is available through help pages, which describe inputs, outputs, and intended use of individual functions, and support day-to-day analytical work. In addition, narrative documentation is provided through vignettes, which illustrate typical analytical workflows and practical use cases. Together, these materials support onboarding of new team members and promote consistent application of shared analytical practices across projects.

## Contribution and validation process

Beyond the technical implementation, **grstat** is embedded within a formal contribution and validation process designed to ensure methodological consistency, software reliability, and progressive adoption in real-world analyses.

When a new need is identified, such as a feature request, an improvement, or a bug report, it is formally documented together with relevant methodological and technical considerations. The request is then reviewed by the working group, which assigns responsibility for implementation. Development takes place in an isolated environment, allowing experimentation without affecting the main codebase. Each proposed contribution must include the R code implementation itself, comprehensive documentation, and a set of automated tests.

Automated testing plays a central role in the validation process. Tests consist of targeted checks designed to verify expected function behaviour and to detect unintended changes over time. This approach supports code robustness and backward compatibility as the package evolves. Each contribution is reviewed by at least one experienced member of the group to assess both code quality and methodological soundness before integration into the main codebase.

New functions are initially released as experimental, with warning messages displayed at runtime to encourage users to critically assess the outputs. They are then applied in real-life projects, and structured feedback is collected from users within the unit. Once sufficient practical experience has been accumulated and any necessary refinements have been made, the feature can be promoted to stable status and considered production-ready.

Overall, the feature validation process follows a three-step model:

- **Development phase**: The feature is under active development. Its code is isolated in a separate work environment to allow safe experimentation and joint technical and methodological review.
- **Beta phase**: Once the feature is integrated into the main codebase, it becomes available to users. Calling the function triggers a warning to indicate that it is still experimental.
- **Stable phase**: The feature is considered stable after receiving sufficient positive feedback from real-world use. At this stage, efforts are made to ensure backward compatibility.

Of note, a specific situation arose during the translation of existing, previously validated SAS macros into R functions. In this case, the validation process involved a systematic comparison of outputs between the SAS and R implementations, using both real-life and simulated datasets. This was made possible through a SAS–Python interface (saspy) and a Python–R bridge (reticulate) [1,2], allowing end-to-end reproducibility across languages.

# Technical implementation

This section describes the practical implementation of our system, relying on standard tools from the R and GitHub ecosystems.

## Issue Tracking

Needs are documented through GitHub issues, which serve as the central entry point for requests, discussions, and decision-making. Issues allow contributors to clarify expected behaviour, suggest alternatives, share examples, and discuss implementation details asynchronously. They also provide structured metadata, including labels, milestones, priority tags, and links to related code or previous discussions, ensuring traceability over time.

## Code governance and review process

Changes are implemented in dedicated development environments in the form of feature git branches on GitHub, allowing contributors to develop or modify functions without affecting the main codebase. Completed work is submitted for review via a Pull Request, which summarises the proposed changes and enables discussion, feedback, and revisions. Once approved, the Pull Request is merged into the main branch, making the new or updated feature available to all users. The main branch is protected and cannot be modified directly, ensuring that all changes undergo review and approval before integration.

## Documentation

Function-level documentation is generated using the R package **roxygen2** [3], while workflow-oriented explanations are provided through vignettes written in R Markdown. These materials support both day-to-day use and training for complete analytical workflows.

## Testing

Automated tests are written using the R package **testthat** [4] and executed locally during development. The full test suite is also run automatically for every Pull Request via GitHub Actions, ensuring that updates cannot introduce regressions and that the package remains backward-compatible.

## GitHub Actions: Continuous integration and deployment

GitHub Actions is a free automation service integrated into GitHub. In this project, GitHub Actions automates several tasks, including running the test suite, building the documentation website with the R package **pkgdown** [5], and managing versioning to maintain clear traceability throughout the project.

# Package Features

The package provides a growing set of functions supporting routine analytical tasks, mainly for oncology clinical trials, with features released at different stages of validation.

Functions reaching the stable phase currently include only tools for adverse event summarisation and visualisation, which are directly derived from previously validated SAS macros used within the unit. Several analytical and reporting features are currently available in a beta phase, including visualisation tools for treatment response, templates for standardised reporting, and functions supporting specific trial designs. Further functionalities, such as standardised Cox model reporting and randomisation list generation, are under active development.

Additional utilities that do not require formal statistical validation are also included, with one function for generating standardised project structures and another for producing a simulated clinical trial dataset used consistently across examples, documentation, and tests.

The **grstat** package builds on other widely used R packages, including components of the the tidyverse (particularly dplyr, ggplot2, and tidyr), as well as cli, crosstable, and flextable [6–9]. Its development relies on framework packages such as devtools, usethis, testthat, and roxygen2 [3,4,10,11]. The complete list of dependencies can be found in the DESCRIPTION file [12].

## Benefit

As outlined throughout this article, the shared toolbox supports greater efficiency and consistency by centralising common biostatistical procedures, hence reducing duplication and occurrence of errors. Its design aims to benefit statisticians with heterogeneous experience in R programming.

The transition from legacy SAS macros to a modular R package enabled the adoption of contemporary software development practices, including version control with Git, web-based documentation in place of static local files, and continuous integration with automated testing. Together, these practices contribute to improved traceability, maintainability, and reliability of the analytical codebase.

Code review plays a central role in this framework by promoting collaboration and knowledge sharing, facilitating early detection of errors, and supporting collective ownership of the code. It also contributes to skill development within the team by providing structured feedback and reinforcing methodological consistency across analyses.

Overall, this approach strengthens the robustness and transparency of biostatistical workflows within the unit, supporting more reliable analytical outputs in the context of clinical research.

## Limitations

The package has several limitations. It relies on external dependencies, which may introduce compatibility issues when upstream packages change. Some functions embed methodological choices specific to our Clinical Trial Unit, which assumptions may not generalise to all contexts. Even when clearly documented, they require careful consideration, as validating such assumptions calls for expertise that automated tests cannot replace. Adoption still depends on staff motivation to use R, which may be limited by the effort required to transfer existing SAS programming skills, although the package should help ease this transition. Regarding tests, some edge cases may still need to be explored, and the test coverage (i.e. the proportion of lines of code executed by at least one test) needs improvement. Although the documentation is expanding, it may not yet cover all use cases. Finally, the development and maintenance of the package require sustained resources, which may constrain the speed of future enhancements.

## Future Scope

As the package becomes more widely integrated into routine workflows within the unit, sustained use and user feedback will contribute to increased robustness and maturity of the implemented functions. Practical experience will progressively highlight limitations, unmet needs, and opportunities for refinement, supporting the development of a toolbox that can evolve alongside methodological advances in biostatistics and changing project requirements. This iterative process is expected to further improve efficiency, reliability, and reproducibility across studies.

In a second phase, the tutorial vignettes will be migrated to a dedicated companion project structured as a book. This resource will provide a more coherent and accessible framework for training and onboarding, allowing both new and experienced staff to consolidate their understanding of core concepts, methodological choices, and recommended practices that extend beyond the package's direct scope.

We hope that this experience report may serve as a source of inspiration for other clinical research units facing similar organisational and methodological challenges.

# References

1. saspy: A Python interface to SAS [Internet]. [cited 2026 Jan 16]. Available from: https://github.com/sassoftware/saspy

2. Kalinowski T, Ushey K, Allaire JJ, RStudio, Tang [aut Y, cph, et al. reticulate: Interface to "Python" [Internet]. 2025 [cited 2026 Jan 16]. Available from: https://cran.r-project.org/web/packages/reticulate/

3. Wickham H, Danenberg P, Csárdi G, Eugster M, Software P, PBC [cph, et al. roxygen2: In-Line Documentation for R [Internet]. 2025 [cited 2026 Jan 16]. Available from: https://cran.r-project.org/web/packages/roxygen2

4. Wickham H, Software P, PBC, utils::recover()) RC team (Implementation of. testthat: Unit Testing for R [Internet]. 2026 [cited 2026 Jan 16]. Available from: https://cran.r-project.org/web/packages/testthat/

5. Wickham H, Hesselberth J, Salmon M, Roy O, Brüggemann S, Software P, et al. pkgdown: Make Static HTML Documentation for a Package [Internet]. 2025 [cited 2026 Jan 16]. Available from: https://cran.r-project.org/web/packages/pkgdown/

6. Wickham H, RStudio. tidyverse: Easily Install and Load the "Tidyverse" [Internet]. 2023 [cited 2026 Jan 16]. Available from: https://cran.r-project.org/web/packages/tidyverse/

7. Csárdi G, Wickham H, Müller K, Brüggemann S, Software P, PBC. cli: Helpers for Developing Command Line Interfaces [Internet]. 2025 [cited 2026 Jan 16]. Available from: https://cran.r-project.org/web/packages/cli/

8. Chaltiel D, Hajage D. crosstable: Crosstables for Descriptive Analyses [Internet]. 2025 [cited 2026 Jan 16]. Available from: https://cran.r-project.org/web/packages/crosstable/

9. Gohel D, ArData, Jager C, Daniels E, Skintzos P, Fazilleau Q, et al. flextable: Functions for Tabular Reporting [Internet]. 2025 [cited 2026 Jan 16]. Available from: https://cran.r-project.org/web/packages/flextable/

10. Wickham H, Hester J, Chang W, Bryan J, Software P, PBC [cph, et al. devtools: Tools to Make Developing R Packages Easier [Internet]. 2025 [cited 2026 Jan 16]. Available from: https://cran.r-project.org/web/packages/devtools/

11. Wickham H, Bryan J, Barrett M, Teucher A, Software P, PBC [cph, et al. usethis: Automate Package and Project Setup [Internet]. 2025 [cited 2026 Jan 16]. Available from: https://cran.r-project.org/web/packages/usethis/

12. grstat package, DESCRIPTION file [Internet]. GitHub. [cited 2026 Jan 16]. Available from: https://github.com/Oncostat/grstat/blob/main/DESCRIPTION